\documentclass[lettersize,journal]{IEEEtran}
\usepackage{amsfonts}
\usepackage{amsmath}
\usepackage{amssymb}
\usepackage{algorithmic}
\usepackage{algorithm}
\usepackage{array}
\usepackage{bm}
\usepackage{textcomp}
\usepackage{stfloats}
\usepackage{multirow}
\usepackage{footnote}
\usepackage{url}
\usepackage{verbatim}
\usepackage{graphicx}
\usepackage{float}
\usepackage{subfigure}
\usepackage{cite}
\usepackage[colorlinks,citecolor=blue,linkcolor=blue,urlcolor=blue]{hyperref}
\usepackage{booktabs}
\usepackage{xcolor}
\usepackage{colortbl}
\def\BibTeX{{\rm B\kern-.05em{\sc i\kern-.025em b}\kern-.08em
		T\kern-.1667em\lower.7ex\hbox{E}\kern-.125emX}}
\usepackage{balance}
\begin{document}
	\title{Intelligent Sky Mirrors: SAC-Driven MF-RIS Optimization for Secure NOMA in Low-Altitude Economy}
	\author{Sai Zhao, Fanjin Kong,  Dong Tang, Tuo Wu, Shunxing Yang,\\  Kai-Kit Wong, \emph{Fellow}, \emph{IEEE}, Kin-Fai Tong, \emph{Fellow, IEEE},\\ and Kwai-Man Luk, \IEEEmembership{Life Fellow,~IEEE}
		\\   
		\thanks{(\textit{Corresponding author: Tuo Wu.}) }
		\thanks{S. Zhao, F. Kong and D. Tang are with the School of Electronics and Communication Engineering, Guangzhou University, Guangzhou 510006, China (E-mail: $\rm \{zhaosai, tangdong\}@gzhu.edu.cn, fanjin@e.gzhu.edu.cn$). T. Wu and K.-M. Luk  are with the State Key Laboratory of Terahertz and Millimeter Waves, Department of Electronic Engineering, City University of Hong Kong, Hong Kong (E-mail: $\rm \{tuowu2, eekmluk\}@cityu.edu.hk$). S. Yang is with China Telecom Research Institute, Guangzhou 510000, China (E-mail: $\rm yangsx@chinatelecom.cn$). K.-K. Wong is with the Department of Electronic and Electrical Engineering, University College London, WC1E 6BT London, U.K., and also with the Yonsei Frontier Laboratory and the School of Integrated Technology, Yonsei University, Seoul 03722, South Korea (E-mail:$\rm  kai$-$\rm kit.wong@ucl.ac.uk$).	  K. F. Tong is with the School of Science and Technology, Hong Kong Metropolitan University, Hong Kong SAR, China. (E-mail: $\rm \{byliu,ktong\}@hkmu.edu.hk$).   This research work of T. Wu was funded by Hong Kong Research Grants Council under the Area of Excellence Scheme under Grant AoE/E-101/23-N.  } 
	}
	

	\maketitle
	
	\begin{abstract}
		Low-altitude economy (LAE) has become a key driving force for smart cities and economic growth. To address spectral efficiency and communication security challenges in LAE, this paper investigates secure energy efficiency (SEE) maximization using intelligent sky mirrors, UAV-mounted multifunctional reconfigurable intelligent surfaces (MF-RIS) assisting nonorthogonal multiple access (NOMA) systems. These aerial mirrors intelligently amplify legitimate signals while simultaneously generating jamming against eavesdroppers. We formulate a joint optimization problem encompassing UAV trajectory, base station power allocation, RIS phase shifts, amplification factors, and scheduling matrices. Given the fractional SEE objective and dynamic UAV scenarios, we propose a two-layer optimization scheme: SAC-driven first layer for trajectory and power management, and channel alignment-based second layer for phase optimization. Simulations demonstrate that our proposed scheme significantly outperforms benchmark approaches.
	\end{abstract}
	

	\section{Introduction}
	Low-altitude economy (LAE)  has become a key driving force for the development of smart cities and economic growth. With the rapid expansion of the Internet of Things (IoT), urban air mobility, and emergency rescue operations, LAE places strict demands on the coverage and performance of communication networks\cite{add_3}\cite{add_4}. This driving force has prompted the accelerated evolution of communication technologies, aiming to enhance spectral efficiency, optimize energy consumption, and ensure communication security and reliability  \cite{1,add_5,add_6,add_7}.
	
	In this context, reconfigurable intelligent surfaces (RIS) mounted on unmanned aerial vehicles (UAVs) emerge as intelligent sky mirrors for low-altitude communications. These aerial mirrors dynamically adjust signal phases and strategically position themselves to optimize propagation in complex environments. Existing research has explored this synergy: Yang et al. \cite{2} jointly optimized beamforming, artificial noise, UAV positioning, and RIS phase shifts to maximize worst-case secrecy rates. Another study \cite{3} enhanced wireless security through joint optimization of virtual partitioning and three-dimensional aerial RIS deployment. Moreover, \cite{4} demonstrated that energy management is crucial for sustainable LAE development. 
	
	To meet increasing user access demands in LAE, nonorthogonal multiple access (NOMA) technology has gained attention for improving spectral efficiency and supporting large-scale connections \cite{5}. However, NOMA systems face security challenges as eavesdroppers can intercept superimposed user signals. Traditional RIS and MF-RIS with amplification, reflection, and refraction struggle to address this challenge \cite{add_1,add_2}. Therefore, we employ a novel MF-RIS that enhances physical layer security by intelligently assigning dual functions to active elements: generating jamming signals while simultaneously strengthening legitimate transmissions. Previous studies \cite{6,7} investigated fixed MF-RIS for NOMA security, while \cite{8} explored UAV-mounted MF-RIS for single-user scenarios but overlooked multi-user cases and energy considerations. Therefore, integrating these intelligent aerial mirrors with NOMA systems provides a promising solution for the security-efficiency trade-off in LAE. 
	
	This paper addresses LAE communication challenges by optimizing secure energy efficiency (SEE) through intelligent sky mirrors in NOMA systems. We jointly optimize UAV trajectory, base station power allocation, RIS amplification factors, phase shifts, and scheduling matrices to maximize SEE. The fractional SEE objective and dynamic aerial mirror mobility necessitate a novel two-layer optimization framework: SAC-driven first layer for trajectory and power management, and channel alignment-based second layer for phase optimization. This approach effectively handles problem non-convexity and time-varying channels. Simulations demonstrate that our proposed scheme significantly outperforms benchmarks, validating the effectiveness of SAC-driven intelligent sky mirrors for secure and energy-efficient LAE communications. 
	
	\begin{figure}[!t]
		\centering
		\includegraphics[width=0.8\columnwidth]{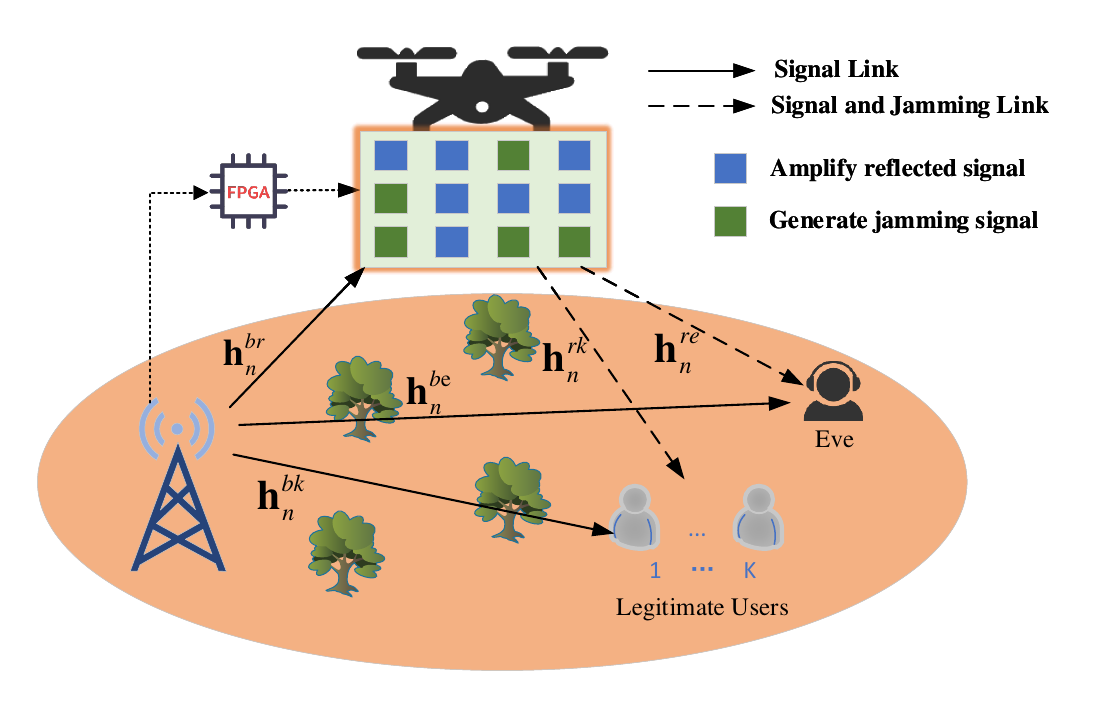}
		\caption{Intelligent sky mirrors assisted NOMA downlink system.}
		\label{fig1}
	\end{figure}
	
	\section{System Model and Problem Formulation}
	
	\subsection{System Model}
	As shown in Fig. \ref{fig1}, we consider intelligent sky mirrors, UAV-equipped MF-RIS (MR-UAV), assisted NOMA downlink system. Both the BS and ground users $g \in G =\{\textrm{Eve}, 1, \ldots,  K\} $ (including $K$ legitimate users and an eavesdropper (Eve)) are equipped with single antenna.
	
	The total time $N$ consists of $n$ time slots of length $\delta$. The UAV flies at a fixed altitude $H$ to avoid collisions with obstacles \cite{9}, and its horizontal coordinates at time slot $n$ are denoted as $L_{n}^{u}=(x_{n}, y_{n})$. In each time slot, the mobility strategy of the UAV consists of the horizontal flight direction $\theta_{n} \in[0,2\pi)$ and the horizontal flight distance $m_{n} \in[0, m_{\max}]$. Accordingly, the flight energy consumption of the UAV in time slot $n$ is expressed as
	\begin{equation}
		\label{eq:1}
		\begin{split}
			{P}_{n}^{\textrm{uav}}&=\delta\bigg[
			P_0\left(1+\frac{3v_n^2}{U_{\textrm{tip}}^2}\right)+\frac{1}{2}d_0\rho sGv_n^3\\
			&\quad+P_1\left(\sqrt{1+\frac{v_n^4}{4v_0^4}}-\frac{v_n^2}{2v_0^2}\right)^{1/2} \bigg]
		\end{split}
	\end{equation}
	where $v_{n}=m_{n} / \delta$ is the horizontal flight velocity; $P_{0}$ and $P_{1}$ are the blade profile power and induced power during hovering, respectively; $U_{\textrm{tip}}$ and $v_0$ denote the rotor blade tip speed and the mean rotor-induced velocity during hover, respectively; $d_0$ is the fuselage drag ratio; $s$ is rotor solidity; $\rho$ and $G$ denote the air density and rotor disk area, respectively \cite{10}.
	
	The channel coefficients between BS and RIS, RIS and the ground user, and BS and the ground user are denoted by $\mathbf{h}^{br}_n=[h^{br}_{1},\ldots,h^{br}_{m},\ldots,h^{br}_{M}]^T\in \mathbb{C}^{M\times1}$, $\mathbf{h}^{rg}_n=[h^{rg}_{1},\ldots,h^{rg}_{m},\ldots,h^{rg}_{M}]\in \mathbb{C}^{1\times M}$, and ${h}^{bg}_n\in \mathbb{C}^{1\times1}$, respectively. These channels are modeled as follows \cite{11}
	\begin{align}
		{h}^{br}_m &= \sqrt{\frac{h_0}{d^{\kappa_1}}} h^{br}_{m,\textrm{LoS}}, \quad m\in M,\nonumber\\
		{h}^{rg}_m &= \sqrt{\frac{h_0}{d^{\kappa_2}}} 
		h^{rg}_{m,\textrm{LoS}}, \quad m\in M, g\in G, \nonumber\\
		{h}^{bg}_n &= \sqrt{\frac{h_0}{d^{\kappa_3}}}  {h}_{\textrm{NLoS}}, \quad g\in G.
	\end{align}
	
	\noindent where $h_0$, $d$, $\kappa$ represent the path loss, Euclidean distance, and the path loss exponent, respectively. $h^{br}_{m,\textrm{LoS}}$ and $h^{rg}_{m,\textrm{LoS}}$ are the line-of-sight (LoS) components. ${h}_{\textrm{NLoS}}$ is the non-line-of-sight (NLoS) component.
	
	The RIS element phase shift $\theta^{m,g}_n$ is designed to enhance the gain of the combined channel ${h}_n^g={h}_m^{rg}\theta^{m,g}_n{h}_m^{br}+{h}_n^{bg}$ by compensating for the cascaded channel phase and aligning the cascaded channel phase with the direct channel phase (i.e., $\angle\left({{h}_m^{rg}\theta^{m,g}_n{h}_m^{br}}\right)=\angle{h}_n^{bg}$, where $\angle\cdot$ denotes the phase of complex numbers) \cite{12,13}. In this paper, we consider optimal RIS phases that flexibly align the RIS elements with ground users. 
	
	The RIS element scheduling matrix is denoted by $\mathbf{A}^g_n=\textrm{diag}(\alpha^{1,g}_n,\ldots,\alpha^{M,g}_n)$, where $\alpha^{m,g}_n\in\{0,1\}$ is the channel alignment indicator for the $m$-th element with the $g$-th ground user. Moreover, $\mathbf{\vartheta}^g_n=\textrm{diag}(\theta^{1,g}_n,\ldots,\theta^{M,g}_n)$ and $\boldsymbol{\beta}_n=\textrm{diag}(\sqrt{\beta^1_n},\ldots,\sqrt{\beta^M_n})$ denote the phase shift matrix of the RIS aligned with the $g$-th ground user and the amplification factor matrix of the RIS, respectively. 
	
	We assume that MR-UAV has the functions of simultaneously amplifying the reflected signal and generating jamming signals \cite{6,7}. Specifically, the overall phase shift matrix $\mathbf{\Theta}_n$ at the RIS consists of two parts: $\mathbf{\Theta}_n=\mathbf{\Theta}^R_n+\mathbf{\Theta}^J_n$, where $\mathbf{\Theta}^R_n=\sum_{k=1}^K\mathbf{A}^k_n\mathbf{\vartheta}^k_n\boldsymbol{\beta}_n$ and $\mathbf{\Theta}^J_n=\mathbf{A}_n^e\mathbf{\vartheta}_n^e\boldsymbol{\beta}_n\boldsymbol{\Phi}$. Here, the modulation matrix $\boldsymbol{\Phi}$ is used to confuse Eve by changing the phase of the incident signal irregularly and dynamically over different symbol periods. Since $\boldsymbol{\Phi}$ only affects the jamming generation process, while $\mathbf{\vartheta}^g_n$ is more relevant to the system performance, we ignore the design of $\boldsymbol{\Phi}$ here and focus more on the effect of $\mathbf{\vartheta}^g_n$ on the performance \cite{6,7}.

	\subsection{Signal and Channel Model}
	Based on the above channel modeling and RIS design, the signal transmitted from the BS is expressed as
	\begin{equation}
		\mathbf{x}_n=\sum_{k=1}^K p^k_n s^k_n.
	\end{equation}
	
	\noindent where $p^k_n$ is the transmit power of user $k$ in time slot $n$, and $\mathbb{E}\{|s^k_n |^2\}=1$. Therefore, the received signal at the $g$-th ground user in time slot $n$ is formulated as 
	\begin{equation}
		{y}^g_n={h}^g_n\mathbf{x}_n+\mathbf{h}^{rg}_n\mathbf{z}_n+\mathbf{h}_n^{rg}\mathbf{\Theta}_n n_s+n_0.
	\end{equation}
	
	\noindent where ${h}_n^g=\mathbf{h}_n^{rg}\mathbf{\Theta}_n^R\mathbf{h}_n^{br}+{h}_n^{bg}$ represents the combined channel, and $\mathbf{z}_n=\mathbf{\Theta}_n^J\mathbf{h}_n^{br}\mathbf{x}_n$ is jamming signal generated at RIS. Furthermore, $n_s$ and $n_0$ denote the thermal noise with power $\sigma_1^2$ and the AWGN with power $\sigma_0^2$, respectively.
	
	\subsection{NOMA and Security Considerations}
	Without loss of generality, we arrange the decoding order according to the channel gain \cite{5}, i.e., $|{h}_n^{i}|^2\geq\ldots\geq|{h}_n^{j}|^2$. Therefore, the SINR for user $i$ decoding user $j$ is denoted as
	\begin{align}
		\gamma_n^{i, j}&=\frac{|{h}_n^{i}|^2p_n^j}{\sum_{l=j+1}^K p_n^l|{h}_n^i|^2+|\mathbf{h}_n^{ri}\mathbf{z}_n|^2+\sigma_1^2\|\mathbf{h}_n^{ri}\mathbf{\Theta}_n\|^2+\sigma_0^2}, \nonumber\\
		&j=\{1,\ldots,K \}, i=\{j+1,\ldots,K \}.
	\end{align}
	
	According to NOMA principle, the target SINR of user $j$ must be less than or equal to the decoded SINR of other stronger users, i.e., the constraint $ \gamma_n^{j,j}\leq\gamma_n^{i,j}$ should be satisfied. The target rate at user $j$ is expressed as \cite{14}
	\begin{align}
		R_n^{j}&=\log_2(1+\min(\gamma_n^{j,j},\gamma_n^{j+1,j},\ldots,\gamma_n^{K,j})),\nonumber\\
		&j=\{1,\ldots,K \}.\label{eq:6}
	\end{align}
	
	In the worst-case scenario, assuming Eve has the ability to eliminate inter-user interference, the eavesdropping rate for the Eve to wiretap user $j$ can be expressed as
	\begin{equation}
		\label{eq:7}
		R_n^{e,j}=\log_2\left(1+\frac{|{h}_n^e|^2p_n^j}{|\mathbf{h}_n^{re}\mathbf{z}_n|^2+\sigma_1^2\|\mathbf{h}_n^{re}\mathbf{\Theta}_n\|^2+\sigma_0^2}\right).
	\end{equation}
	
	\indent Combining (\ref{eq:1}), (\ref{eq:6}), and (\ref{eq:7}), the SEE is expressed as
	\begin{equation}
		\textrm{SEE}_n=\frac{\sum_{k=1}^{K}[R_n^k-R_n^{e,k}]^+}{{P}_n^{\textrm{sum}}={P}_n^{\textrm{ris}}+{P}_n^{\textrm{uav}}}.
	\end{equation}
	
	\noindent where ${P}_n^{\textrm{ris}}=\sum_{k=1}^{K}p_n^k\|\mathbf{\Theta}_n\mathbf{h}_n^{br}\|^2+\sigma_1^2\|\mathbf{\Theta}_n\|_F^2$ is the power consumption of RIS.
	
	\subsection{Problem Formulation}
	In the emerging LAE, the large-scale application of UAV puts severe demands on communication security and the management of UAV energy consumption. Therefore, our objective is to maximize the SEE of the MR-UAV-assisted NOMA downlink system by jointly optimizing the power allocation $\mathbf{P_n}=\{p^1_n, \ldots,p^K_n\}$, RIS element scheduling matrix $\mathbf{A}^g_n$, RIS phase shift matrix $\mathbf{\vartheta}^g_n$, RIS amplification factor matrix $\boldsymbol{\beta}_n$, UAV flight distance $m_{n}$, and UAV horizontal flight angle $\theta_n$. The optimization problem is formulated as
	
	\begin{subequations}\label{eq:opt}
		\begin{align}
			(\mathbf{P}_0)\quad&\max_{\{\mathbf{P}_n,\mathbf{A}^g_n,\mathbf{\vartheta}^g_n,\boldsymbol{\beta}_n,m_n,\theta_n\}}\quad\sum_{n=1}^N \textrm{SEE}_n\quad \label{eq:opt_a}\\
			&\textrm{s.t.}\quad\sum_{k=1}^K p_n^k\leq P_{b,\max},\quad\forall n \in \mathcal{N}, \label{eq:opt_b} \\
			&\quad\quad\;\; {P}_n^{\textrm{ris}}\leq P_{r,\max},\quad\forall n \in \mathcal{N}, \label{eq:opt_c}\\
			&\quad\quad\;\; R_n^k\geq Q_{\min},\quad\forall k \in \mathcal{K}, \forall n \in \mathcal{N}, \label{eq:opt_d}\\
			&\quad\quad\;\; \theta^{m,g}_n \in[0,2\pi),\quad\forall m\in \mathcal{M}, \forall g\in G, \forall n \in \mathcal{N}, \label{eq:opt_e}\\
			&\quad\quad\;\; L_{n}^u\in\mathcal{L}^{\textrm{fly}}, \quad\forall n \in \mathcal{N}, \label{eq:opt_f}
		\end{align}
	\end{subequations}
	
	\noindent where $\mathcal{N}=\{1,2,\ldots,N\}$, $\mathcal{K}=\{1,2,\ldots,K\}$, $\mathcal{M}=\{1,2,\ldots,M\}$, and ${G}=\{\textrm{Eve}, 1, \ldots, K\}$ represent the sets of time slots, users, RIS elements, and ground nodes, respectively. Specifically, \eqref{eq:opt_b} and \eqref{eq:opt_c} are the BS transmit power constraint and the RIS output power constraint, respectively. \eqref{eq:opt_d} represents the user's quality of service (QoS) constraint. \eqref{eq:opt_e} is the RIS phase shift constraint, and \eqref{eq:opt_f} is the flight space constraint. Due to the fractional form of SEE and the dynamic scenario of the UAV, it is difficult for traditional optimization methods to solve this problem effectively. Therefore, we propose a two-layer optimization scheme  with low complexity and faster convergence based on the SAC algorithm and the channel alignment method.

	\section{Proposed Two-Layer Optimization Scheme}
	
	Due to the fractional form of SEE and the dynamic nature of the UAV-assisted system, Problem $\mathbf{P}_0$ is a non-convex mixed-integer optimization problem that is challenging to solve using traditional optimization methods. To address this challenge, we propose a two-layer optimization scheme that decomposes the problem into manageable subproblems while maintaining computational efficiency.
	
	In the proposed scheme, the first layer employs the SAC algorithm to jointly optimize the UAV trajectory, BS transmit power allocation, RIS amplification factor matrix, and RIS element scheduling matrix. The second layer utilizes a channel alignment method to calculate the optimal phase shift of RIS elements. This decomposition strategy reduces computational complexity while ensuring fast convergence.
	
	\subsection{SAC-Based First Layer Optimization}
	The Soft Actor-Critic (SAC) algorithm is an off-policy actor-critic method that effectively addresses the exploration-exploitation trade-off in deep reinforcement learning by maximizing both the expected reward and the policy entropy \cite{15}. This dual objective ensures that the learned policy is not only efficient but also sufficiently exploratory, making it particularly suitable for complex optimization problems like $\mathbf{P}_0$.
	
	SAC is well-suited for our problem due to its ability to: (1) efficiently utilize collected experience data through off-policy learning, (2) handle high-dimensional continuous action spaces, and (3) maintain stable learning in dynamic environments. The algorithm operates within a Markov Decision Process (MDP) framework, learning optimal policies through agent-environment interactions.
	
	To formulate our optimization problem within the SAC framework, we define the MDP as a 4-tuple $\langle \mathbb{S},\mathbb{A},\mathbb{R},\mathbb{P}\rangle$, where:
	
	\subsubsection{State Space}
	The state space $\mathbb{S}$ encompasses all relevant system information required for decision-making. At time slot $n$, the state consists of:  Channel state information: $\mathbf{h}_n=\{\mathbf{h}_n^{br},\mathbf{h}_n^{rg},{h}_n^{bg}\}$;  UAV and ground user positions: $L_n=\{L_n^u,L_n^g\}$  and  RIS phase shift matrix: $\mathbf{\vartheta}_n^g$.
	
	Thus, the state space is formally defined as:
	\begin{equation}
		\mathbb{S}=\{\mathbf{h}_n, L_n, \mathbf{\vartheta}_n^g\}.
	\end{equation}
	
	\subsubsection{Action Space}
	The action space $\mathbb{A}$ comprises all control variables that the agent can adjust at each time slot: transmit power allocation $\mathbf{P}_n$, RIS element scheduling matrix $\mathbf{A}_n$, RIS amplification factor matrix $\boldsymbol{\beta}_n$, UAV horizontal flight distance $m_n$, and UAV horizontal flight angle $\theta_n$.
	
	The complete action space is defined as:
	\begin{equation}
		\mathbb{A}=\{\mathbf{P}_n,\mathbf{A}_n, \boldsymbol{\beta}_n, m_n, \theta_n\}.
	\end{equation}
	
	\subsubsection{Action Mapping}
	Since the policy network output uses the $\tanh(\cdot)$ activation function with range $[-1,1]$, an action mapping process is required to convert continuous outputs to discrete scheduling decisions. For the RIS element scheduling matrix $\mathbf{A}_n=[w^1_{\alpha},\ldots,w^M_{\alpha}]$, each element is mapped to the corresponding user assignment $\mathbf{A}^g_n$. 
	
	For example, with $K=2$ users, the mapping rule is:
	\begin{equation}
		{\alpha^{m,g}_n} = \begin{cases}
			1,&\text{if } \frac{2g'-K-3}{K+1}\leq w^m_{\alpha}<\frac{2g'-K-1}{K+1}, \\ 
			0,&\text{otherwise},
		\end{cases}
	\end{equation}
	where $g' \in \{1, 2, 3\}$ represents the index of ground node $g \in \{\textrm{Eve}, 1, 2\}$.

	Additionally, the transmit power actions $\mathbf{P}_n$ are defined as power allocation ratios. To ensure constraint compliance, when the RIS amplification factor $\boldsymbol{\beta}_n$ violates the power constraints, we apply the following adjustment:
	\begin{equation}
		\hat{\boldsymbol{\beta}}_n=\sqrt{\frac{P_{r,\max}}{{P}_n^{\textrm{ris}}}}\boldsymbol{\beta}_n,\quad\text{if } {P}_n^{\textrm{ris}}>P_{r,\max}.
	\end{equation}
	\subsubsection{Reward Function}
	The reward function design is crucial for guiding the agent toward optimal solutions. Our reward function balances two primary objectives: (1) maximizing the SEE, and (2) penalizing constraint violations. The reward function is formulated as:
	\begin{equation}
		r_n=R_n^{\textrm{sum}}(1-w_1\widehat{P}_n^{\textrm{sum}}-w_2U_n)-p,
	\end{equation}
	where $\widehat{P}_n^{\textrm{sum}}=\frac{{P}_n^{\textrm{sum}}-{P}_{\min}}{{P}_{\max}-{P}_{\min}}$ is the normalized total power consumption, $U_n=\frac{1}{K}\sum_{k=1}^K[Q_{\min}-R_n^k]^+$ represents the degree of QoS constraint violation, $w_1, w_2$ are weighting parameters for power and QoS penalties, and $p$ is the penalty for UAV flying outside the restricted area $\mathcal{L}^{\textrm{fly}}$.

	\subsubsection{SAC Training Process}
	The SAC algorithm operates through two alternating phases: policy evaluation and policy improvement. In the policy evaluation phase, the agent interacts with the environment by observing the current state $s_n$, selecting and executing action $a_n$ according to the current policy, receiving reward $r_n$, and transitioning to the next state $s_{n+1}$. Each experience tuple $\{s_n,a_n,r_n,s_{n+1}\}$ is stored in the replay buffer $\mathbf{D}$ for later training. In the policy improvement phase, the neural networks are updated using mini-batches of $B$ experience transitions randomly sampled from $\mathbf{D}$. This process is repeated $I$ times per episode to ensure stable convergence.

	\subsection{Channel Alignment-Based Second Layer Optimization}
	
	The second layer of our optimization scheme focuses on computing the optimal phase shifts for the RIS elements. Due to the UAV's mobility, the phase alignment matrix $\mathbf{\vartheta}^{g}_n$ must be updated dynamically in each time slot to maintain optimal channel alignment.
	
	The optimal phase shift $\theta^{m,g}_n$ for the $m$-th RIS element aligned with the $g$-th ground user is determined by maximizing the combined channel gain. This is achieved through phase alignment that compensates for the cascaded channel phase:
	\begin{equation}
		\theta^{m,g}_n= e^{j\angle h^{bg}_n}e^{-j\angle h^{br}_m}e^{-j\angle h^{rg}_m}\label{eq:phase_align},
	\end{equation}
	where the phase shifts are computed to align the cascaded channel $h^{br}_m \cdot h^{rg}_m$ with the direct channel $h^{bg}_n$.
	
\begin{figure}[!t]
		\centering
		\includegraphics[width=0.8\columnwidth]{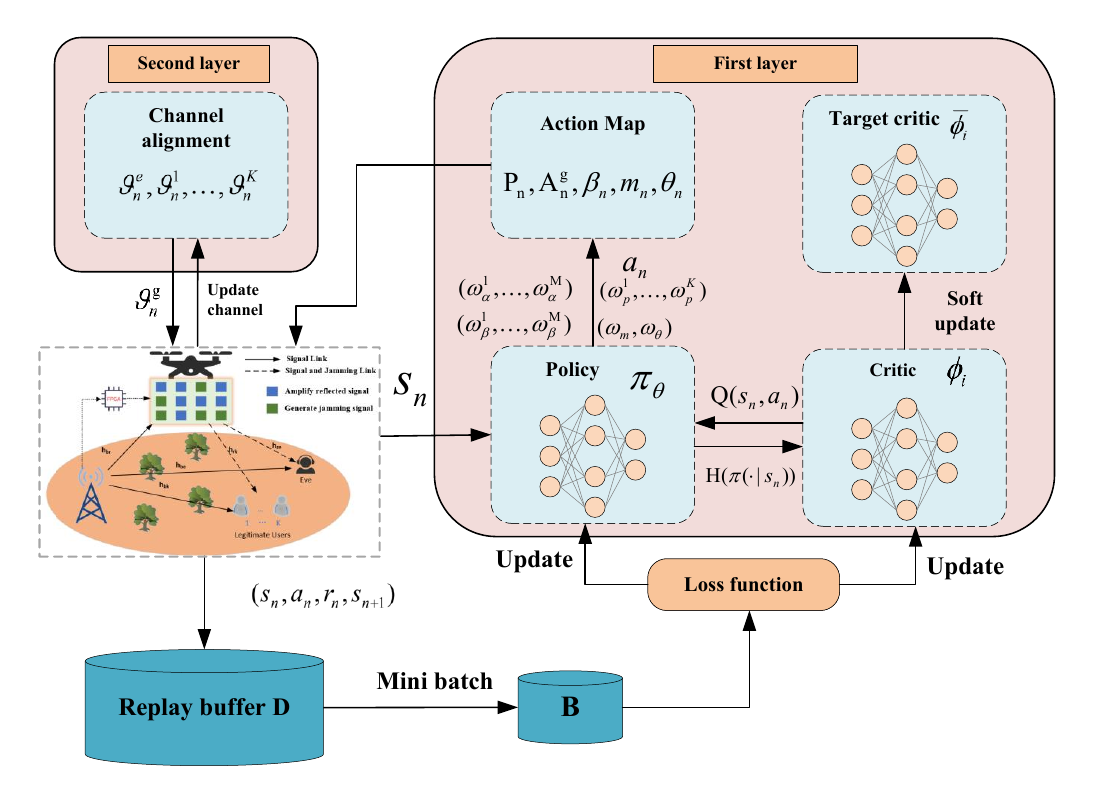}
		\caption{The proposed two-layer optimization structure.}
		\label{fig2}
	\end{figure}

	The phase alignment matrix $\mathbf{\vartheta}_n^{m,g}$ is updated immediately after the SAC agent executes action $a_n$ (i.e., after updating the UAV position), and before calculating the reward $r_n$. The updated phase matrix becomes part of the next state $s_{n+1}$ for the subsequent decision step.  The complete two-layer optimization scheme is  illustrated in Fig. \ref{fig2}.
	
	\section{Numerical Results and Discussion}
	In this section, we present comprehensive numerical results to evaluate the performance of our proposed intelligent sky mirrors assisted NOMA downlink system. The simulation environment consists of a $400\,\textrm{m} \times 400\,\textrm{m}$ square area containing the BS, MR-UAV, and ground users. We consider $K=2$ legitimate users positioned at coordinates $[250,300,1]$ and $[250,315,1]$, respectively. The BS is located at $[10,10,10]$ while the eavesdropper is positioned at $[300,250,1]$. The MR-UAV starts from initial location $[10,390,50]$. The detailed simulation parameters are provided in Table \ref{tab:table1}.
	\begin{table}[!t]
		\caption{Simulation Parameters \label{tab:table1}}
		\centering
		\begin{tabular}{|c|c|}
			\hline 
			$\mathbf{Parameters}$ & $\mathbf{Settings}$\\
			\hline
			$U_{tip}, d_0, \rho, s$ & 120, 0.6, 1.225, 0.05 \\ 
			\hline 
			$G, v_0, P_0, P_1$ & 0.503, 4.3, 79.86, 88.63 \\ 
			\hline 
			$Q_{\min}$ & 1 bits/s/Hz \\ 
			\hline 
			$\sigma_0^2, \sigma_1^2, P_{b,\max}, P_{r,\max}$  (dBm)& -105, -105, 30, 0\\ 
			\hline 
			$\kappa_1, \kappa_2, \kappa_3, h_0$ & 2, 2, 3.6, $1\times10^{-3}$\\ 
			\hline 
			Batch size $B$, replay buffer size  $|\mathbf{D}|$& 256, $1\times10^6$ \\ 
			\hline 
			Episodes $E$, training iterations $I$& 5000, 70\\ 
			\hline 
			Time slots $N$, slot duration $\delta$ & 200, 1s\\ 
			\hline 
			Learning rate $\ell_r$ & $3\times10^{-4}$ \\ 
			\hline 
			Hidden layers & [512, 256] \\
			\hline 
			Maximum flight distance $m_{\max}$ & 20m \\
			\hline 
		\end{tabular}
	\end{table}
	
\begin{figure*}[t]
		\centering
		\begin{minipage}[b]{0.3\linewidth}
			\centering
			\includegraphics[width=\textwidth]{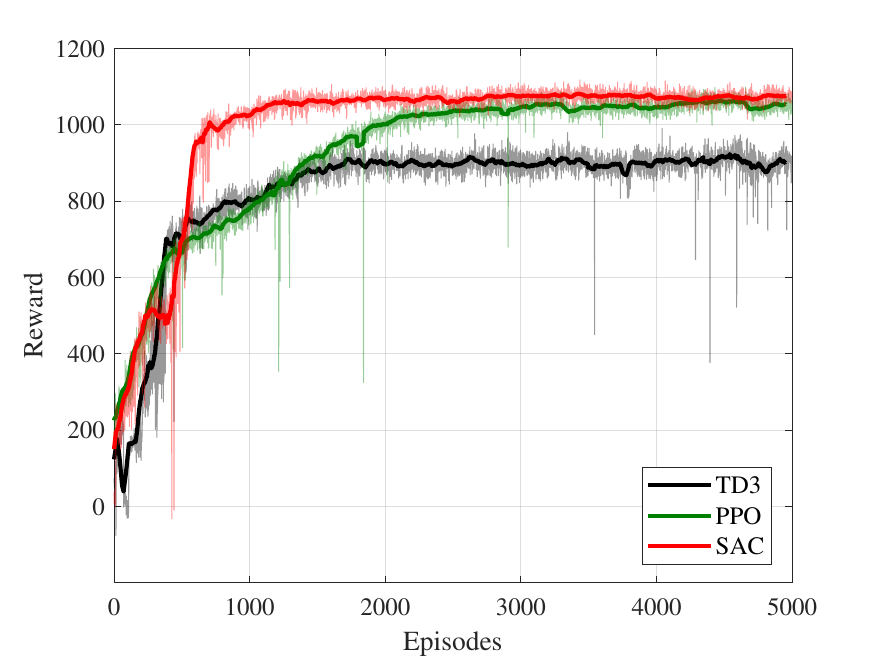}
			\caption{Convergence performance comparison at $M=20$, $\beta_{\max}=8$.}
			\label{fig3}
		\end{minipage}
		\hfill
		\begin{minipage}[b]{0.3\linewidth}
			\centering
			\includegraphics[width=\textwidth]{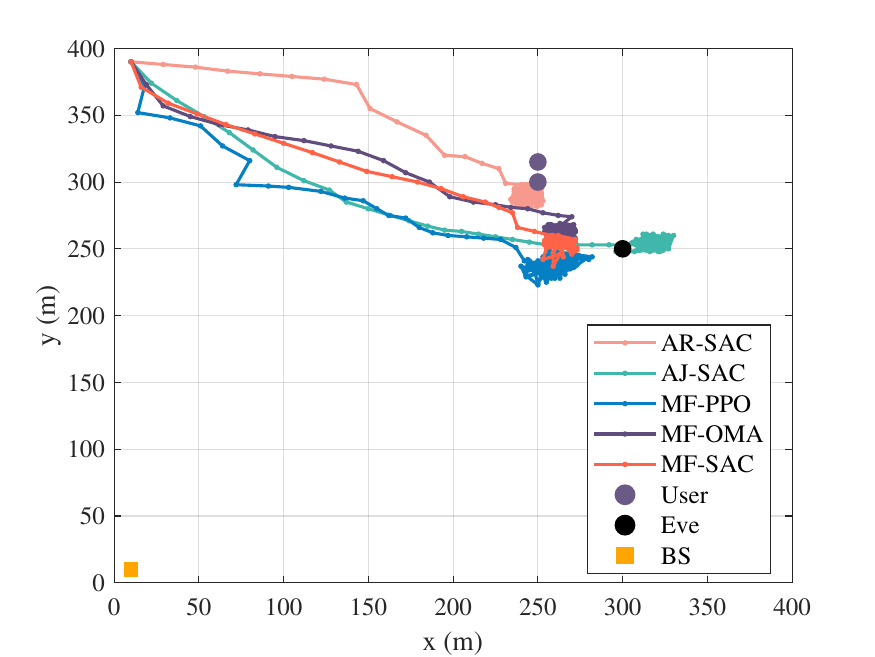}
			\caption{UAV flight trajectory.}
			\label{fig4}
		\end{minipage}
		\hfill
		\begin{minipage}[b]{0.3\linewidth}
			\centering
			\includegraphics[width=\textwidth]{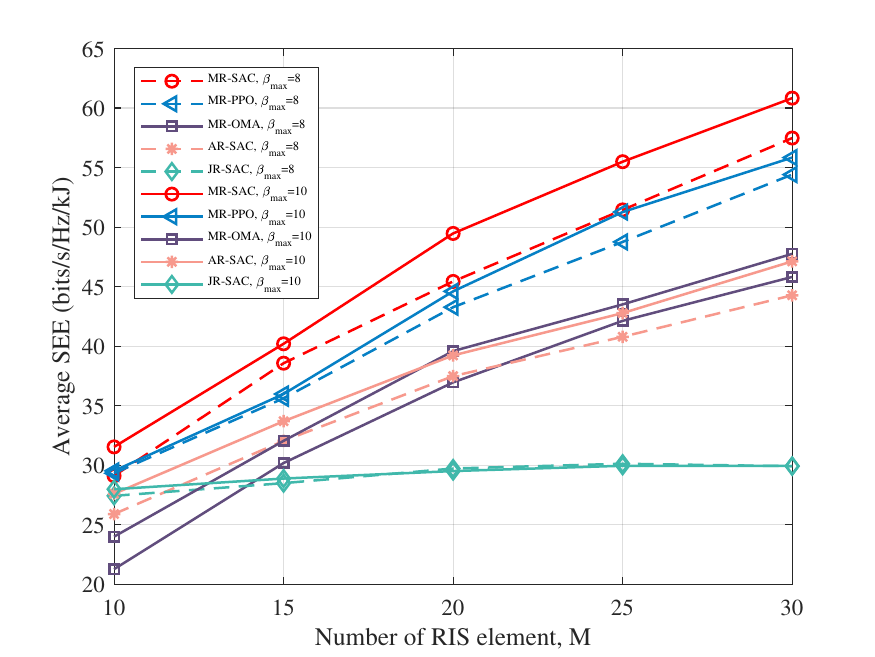}
			\caption{Average SEE versus different number of RIS elements with different $\beta_{\max}$.}
			\label{fig5}
		\end{minipage}
	\end{figure*}

	Fig. \ref{fig3} illustrates the convergence performance comparison among different reinforcement learning algorithms in terms of accumulated reward. We compare our SAC-based approach with two state-of-the-art DRL algorithms: proximal policy optimization (PPO) and twin delayed deep deterministic policy gradient (TD3). The results demonstrate that SAC achieves superior convergence performance with both faster convergence rate and higher reward ceiling compared to PPO and TD3. This superior performance stems from SAC's efficient sample utilization through off-policy learning and its exploration advantage via stochastic policy entropy maximization.

	Fig. \ref{fig4} presents the optimized UAV flight trajectories for all compared schemes, revealing distinct strategic behaviors. All algorithms rapidly converge to their respective optimal hovering positions to minimize energy consumption. The multifunctional schemes (\textbf{MF-SAC}, \textbf{MF-PPO}, and \textbf{MF-OMA}) strategically position the UAV between legitimate users and the eavesdropper, enabling simultaneous enhancement of legitimate communication channels while degrading eavesdropping capability through intelligent signal management. In contrast, \textbf{AR-SAC} (The RIS only actively reflects signals) positions the UAV closer to legitimate users to maximize signal reflection benefits while minimizing information leakage to the eavesdropper. Conversely, \textbf{AJ-SAC} (The RIS only actively generates jamming signals) positions the UAV near the eavesdropper to maximize jamming effectiveness while minimizing interference to legitimate users. Notably, \textbf{MF-SAC} achieves the most efficient trajectory with shorter flight path distance and smoother transitions compared to \textbf{MF-PPO}, demonstrating the superior learning efficiency of the SAC algorithm in UAV trajectory optimization for secure LAE communications.
	
	Fig. \ref{fig5} illustrates the average SEE performance as a function of the number of RIS elements $M$ under different maximum amplification factors $\beta_{\max}$. The results clearly show that increasing the number of RIS elements enhances the SEE for all schemes across different $\beta_{\max}$ values, confirming the benefits of deploying larger intelligent surfaces. Our proposed \textbf{MF-SAC} scheme consistently outperforms all benchmark approaches, achieving the highest SEE values across all configurations. This superior performance validates the effectiveness of the multifunctional RIS design in optimally balancing legitimate signal enhancement and eavesdropper interference, representing a crucial security-efficiency trade-off for LAE applications. Furthermore, comparing results under different amplification settings reveals that higher $\beta_{\max}=10$ consistently yields better SEE than $\beta_{\max}=8$ for all schemes, demonstrating that increased amplification capability directly translates to improved system performance, albeit at potentially higher energy cost.
	
	\section{Conclusion}
	This paper investigated secure energy efficiency maximization using intelligent sky mirrors in LAE-NOMA systems. We formulated a joint optimization problem encompassing UAV trajectory, power allocation, and RIS configurations to maximize SEE. Our novel SAC-driven two-layer optimization framework effectively addresses the fractional SEE objective and dynamic aerial mirror mobility: the first layer employs soft actor critic for trajectory and power management, while the second layer uses channel alignment for phase optimization. Comprehensive simulations validate that our intelligent sky mirrors significantly outperform existing approaches, demonstrating the effectiveness of SAC-driven MF-RIS optimization in achieving superior security-efficiency trade-offs for future LAE communications.

	\bibliographystyle{ieeetr}
	\bibliography{reference}
	
\end{document}